\documentclass{article}
\usepackage{amssymb}
\usepackage{amsmath}

\begin{document}

\title{FORTRAN-codes for an analysis of the ultrashort pulse propagation}
\author{Vladimir L. Kalashnikov \\
Photonics Institute, TU Vienna, Gusshausstr. 27/387,\\
A-1040 Vienna, Austria}
\date{}
\maketitle

\begin{abstract}
Short description of the FORTRAN-codes for an analysis of the
ultrashort pulse dynamics is presented. We consider: 1) the
aberration-less approximation and the momentum method for the
search of the single pulse stability regions in the laser with the
soft-aperture Kerr-lens mode locking; 2) the distributed complex
Ginzburg-Landau model for the same aim; 3) the generalized
Schr\"{o}dinger equation for the analysis of the femtosecond pulse
propagation in the tapered and photonic crystal fibers.
\end{abstract}

\section{Introduction}

Analysis of the intra- and extra-cavity pulse dynamics has the
common features: it can be based on the distributed models of the
complex Ginzburg-Landau type \cite{kramer} or the generalized
Schr\"{o}dinger type \cite{brabec}. The basic difference of
ultrashort pulse laser dynamics from, for example, intra-fiber
propagation of the femtosecond pulses is the essential
contribution of the dissipation into former. However, for the
direct simulation the appearance of the non-Hamiltonian part in
the dynamical equation doesn't cause some problems. It allows
considering the distributed models on the common basis (some
physical and formal backgrounds can be found in
\cite{kalashnikov}).

The main problem results from the transition to the
higher-dimensional models. The trivial dimension is $1+1$, i.e.
propagation distance (or cavity transit number for a laser) $+$
local time (for an analysis of the ultrashort pulse dynamics). Our
experience proved that such dimension is quite sufficient for the
solution of the various problems, for example, spectral continuum
generation in the fibers or multiple pulse operation of the
femtosecond lasers. This approach adequately describes the pulse
dynamics and at the same time is not too complicated to remain
physically meaningful. The minimum set of the governing parameters
allows the overall optimization by
the direct scanning of the parametrical space \cite%
{kalashnikov1,kalashnikov2}.

However, a consideration of the self-start ability of the
soft-aperture Kerr-lens mode-locked lasers (see \cite{brabec}) and
an analysis of the real-world laser configurations requires at the
least $1+2$ dimensions, where one transverse spatial dimension
appears as a result of the rotational symmetry. The direct
simulations on this way exceed the resources of the desk-top
computer, which is common tool for the laser community. Therefore
we have to reduce the dimension, for example, by the means of the
so-called aberration-less approximation. This allows performing
the simulation not on the time-spatial grid but on the grid formed
by the parametrical set defining the trial (aberration-less)
solution.

Below we consider some examples of $1+1$ and $1+2$ dimensional models. The
latter model has in fact reduced dimension due to the use of the momentum
method. It should be noted that the analytical computations underlying the
numerical codes were realized in the MAPLE computer algebra system \cite%
{kalashnikov3}.

\section{Passive mode-locking and nonlinear complex Ginzburg-Landau equation}

This approach is based on 1+1 dimensional model in the framework of the
so-called nonlinear Ginzburg-Landau equation, which describes the Kerr-lens
mode locking as an action of the fast saturable absorber governed by the few
physically meaningful parameters, viz., its modulation depth $\gamma $ and
the inverse saturation intensity $\sigma $.

The master equation describing the ultrashort pulse generation in
the Kerr-lens mode-locked solid-state laser is:

\begin{gather}
\frac{{\partial a(z,t)}}{{\partial z}}=\left[ {\alpha -\rho +t_{f}^{2}\frac{{%
\partial ^{2}}}{\partial t^{2}}-\frac{\gamma }{{1+\sigma \left\vert {a(z,t)}%
\right\vert ^{2}}}}\right] a(z,t)- \\
i\left\{ {\sum\limits_{m=2}^{N}{\frac{{(-i)^{m}\beta _{m}}}{{m!}}\frac{{%
\partial ^{m}}}{{\partial t^{m}}}+}\delta \left( {\left\vert {a(z,t)}%
\right\vert ^{2}-\frac{1}{{\omega _{0}}}\frac{\partial }{{\partial t}}%
\left\vert {a(z,t)}\right\vert ^{2}}\right) }\right\} a(z,t),  \notag
\end{gather}

\noindent where $a(z,t)$ is the field amplitude (so that $|a|^{2}$ has a
dimension of the intensity), $z$ is the longitudinal coordinate normalized
to the cavity length, $t$ is the local time, $\alpha $ is the saturated gain
coefficient, $\rho $ is the linear net-loss coefficient taking into account
the intracavity and output losses, $t_{f}$ is the group delay caused by the
spectral filtering within the cavity, $\beta _{m}$ are the $m$-order
group-delay dispersion coefficients, $\delta $ = $l_{g}n_{2}\omega _{0}/c$ =
$2\pi n_{2}l_{g}/(\lambda _{0}n)$ is the self-phase modulation coefficient, $%
\omega _{0}$ and $\lambda _{0}$ are the frequency and wavelength
corresponding to the minimum spectral loss, $n$ and $n_{2}$ are the linear
and nonlinear refraction coefficients, respectively, $l_{g}$ is the double
length of the gain medium (we suppose that the gain medium gives a main
contribution to the self-phase modulation). The last term in Eq.~(1)
describes the self-steepening effect and for the simplification will be not
taken into account in the simulations. As an additional simplification we
neglect the stimulated Raman scattering in the active medium ~\cite%
{kalashnikov4}.

For the numerical simulations in the framework of the distributed model it
is convenient to normalize the time and the intensity to $t_{f}$ = $\lambda
_{0}^{2}/(\Delta \lambda c)$ and $1/\delta $, respectively ($\Delta \lambda $
is the gain bandwidth). The simulation were performed on the $2^{12}\times
10^{4}$ mesh. Only steady-state pulses were considered. As the criterion of
the steady-state operation we chose the peak intensity change less than 1\%
over last 1000 cavity transits.

Note that the local time interval, which is equal to the cavity period $%
\approx 10$ nanoseconds, is not covered in our case ($2^{12}\times t_{f}$ $%
\approx 20\div 100$ picoseconds). This puts the questions about
stability against the multipulsing with the large inter-pulse
separations. Additionally we can not be sure in the ability of the
spontaneous appearance of the mode locking (problem of the mode
locking self-start ability).

The solution of Eq. (1) is based on the fast Fourier-transform
split-step method (see Appendix 1). We symmetrized non-Hamiltonian
(square brackets in Eq. (1)) and Hamiltonian part (braces in Eq.
(1)) separately. The mode locking in the considered model is
governed by the only four basic parameters: $\alpha -\rho $,
$\beta _{2}$, $\gamma $, and $\sigma $. This allows unambiguous
multiparametric optimization. In the presence of the higher-order
dispersions, the additional $\beta _{m}$ parameters appear. This
complicates the optimization procedure, but keeps its physical
clarity. As an initial condition we take the analytical solution
of the cubic Ginzburg-Landau or Schr\"{o}dinger\ equation
\cite{kalashnikov}.

Some results obtained on the basis of this model are presented in \cite%
{kalashnikov1,kalashnikov2,kalashnikov4}.

\section{Spectral continuum generation in the tapered fiber}

Generation of spectral supercontinuum became a hot topic in optics
in recent years \cite{russell}. In the crystal or tapered fiber
the propagating field has a comparatively determinate spatial
structure due to a strong confinement. Therefore, the
$1+1$-dimensional simulations give a quite thorough result.
However, the nonlinearity in such fibers is enhanced by their
small core size. As a result, a set of the nonlinearities is wider
than that described by the Hamiltonian part of Eq. (1), which is
the high-order nonlinear Schr\"{o}dinger equation. The non-trivial
generalization can be obtain due to taking into account the
stimulated Raman scattering \cite{agrawal}:

\begin{gather}
i\frac{\partial a}{\partial z}+{\sum\limits_{m\geq 2}^{{}}}\frac{i^{m}\beta
_{m}}{m!}\frac{\partial ^{m}a}{\partial t^{m}}{=-\delta }\left( 1-f\right)
\left\vert a\right\vert ^{2}a \\
-\delta f a \int\limits_{-\infty }^{t}R\left( t\right) \left\vert
a\left( z,t-t\right) \right\vert ^{2}dt,  \notag
\end{gather}

\noindent where $\delta =n_{2}\omega _{0}\diagup c$ is the
self-phase modulation coefficient, $\beta _{m}$ is the $m$th-order
group-velocity dispersion coefficient, $f$ is the fraction of the
stimulated Raman
scattering contribution to the nonlinear refractive index of the fiber, $%
R\left( t\right) =\frac{T_{1}^{2}+T_{2}^{2}}{T_{1}^{2}\ T_{2}^{2}}\exp
\left( -\frac{t}{T_{2}}\right) \sin \left( \frac{t}{T_{1}}\right) $ is the
Raman response function \cite{stolen,hollenbeck}. $T_{1}=12.2$ fs and $%
T_{2}=32$ fs define the phonon oscillation period and its dumping time,
respectively.

We normalized $t$ to 1 fs (the normalization to the initial pulse
width is convenient, too), $z$ to the nonlinear length
$L_{nl}=\left( \omega
_{0}n_{2}I_{0}/c\right) ^{-1}$ defined by the initial pulse intensity $I_{0}$%
. The simulations were carried out on the mesh with the time step 1 fs ($%
2^{13}$ points) and the spatial step $10^{-3}L_{nl}$. The solution
of Eq. (2) was based on the fast Fourier-transform split-step
method with the evaluation of the Raman response in the time
domain (see Appendix 2).

The analysis of the pulse propagation in the tapered fibers
requires the attention to the so-called transient sectors: before
(after) the tapered sector with the almost constant waist there is
the convergent (divergent) sector, where the fiber characteristics
change from those in the single-mode fiber to those in the tapered
fiber (or vice versa). The exact law of these changes is unknown,
but we used the simple linear approximation for the evolution of
the intensity and the dispersion coefficients. Note, that the
normalization of the intensity was defined through the parameters
of the tapered sector.

\section{Aberration-less approximation: analysis of the real-world laser
configurations}

Above the problem of the ultrashort pulse stability in the
mode-locked laser were formulated on the basis of the distributed
$1+1$-dimensional Ginzburg-Landau model. We noted also some
problems of such model. Here we present an analysis on the basis
of the time-spatial model. The spatial distribution for the laser
beam is assumed to be Gaussian that reduces the problem to
$1+2$-dimensions. The free-space propagation of the Gaussian beam
can be considered on the basis of the usual ABCD-matrix formalism
\cite{siegman}, while the propagation inside the nonlinear active
medium is described by the following equation:

\begin{gather}
\frac{{\partial a\left( {z,r,t}\right) }}{{\partial z}}-i\left[ {\frac{{%
2\vartheta r^{2}a\left( {z,r,t}\right) }}{{w_{p}^{2}}}-\frac{{\frac{{%
\partial a\left( {z,r,t}\right) }}{{\partial r}}+\frac{{r\partial
^{2}a\left( {z,r,t}\right) }}{{\partial r^{2}}}}}{{2kr}}+\beta _{2}^{\prime
}{}^{2}\frac{{\partial ^{2}a\left( {z,r,t}\right) }}{{\partial t^{2}}}}%
\right] a\left( {z,r,t}\right) \\
+i\chi \left\vert {a\left( {z,r,t}\right) }\right\vert ^{2}a\left( {z,r,t}%
\right) =\alpha \exp \left( {-\frac{{2r^{2}}}{{w_{p}^{2}}}}\right) a\left( {%
z,r,t}\right) +t_{f}^{\prime }{}^{2}\frac{{\partial ^{2}a\left( {z,r,t}%
\right) }}{{\partial t^{2}}}.  \notag
\end{gather}

\noindent Here $\beta _{2}^{\prime }$ and $t_{f}^{\prime }$ are the
group-velocity dispersion and the inverse group-velocity delay coefficients
(for ZnSe laser we used $\beta _{2}^{\prime }{}^{2}$=2054 fs$^{2}$/cm and $%
t_{f}^{\prime }$=13 fs/cm \cite{kalashnikov1}). The left-hand side of Eq.
(3) describes the non-dissipative factors: thermo-lensing ($\vartheta $=$k%
\frac{{dn_{0}}}{{dT}}\zeta P_{a}\exp \left( {-\zeta z}\right) $/ $\left( {%
4\pi n_{0}\kappa _{th}}\right) $, $k$ is the wave number, $\frac{{dn_{0}}}{{%
dT}}$ is the coefficient of the refractive index thermo-dependence (5.35$%
\times 10^{-5}K^{-1}$ for ZnSe), $\zeta $ is the loss coefficient at the
pump wavelength, $P_{a}$ is the pump power, $\kappa _{th}$ is the
thermo-conductivity coefficient (0.172 $WK^{-1}cm^{-1}$ for ZnSe));
diffraction (in the cylindrically symmetrical case); group-velocity
dispersion and self-phase modulation (providing self-focusing for radially
varying beam, $\chi $=$n_{2}k/n_{0}$). The right-hand side of Eq. (3)
describes the dissipative factors inside the gain medium: radially varying
gain (providing gain guiding and soft aperture action, $\alpha $ and $w_{p}$
are the saturated gain coefficient and the pump beam size, respectively);
spectral filtering caused by the gain band profile. The saturated gain can
be expressed in the following way:

\begin{equation}
\alpha =\frac{{2\alpha _{\max }\sigma _{a}P_{a}T_{r}}}{{\hbar \omega _{p}\pi
w_{p}^{2}\left( {\frac{{2\sigma _{a}P_{a}T_{r}}}{{\hbar \omega _{p}\pi
w_{p}^{2}}}+\frac{{2\upsilon P_{g}}}{{\pi w^{2}I_{s}}}\frac{{\tau _{p}}}{{%
T_{cav}}}+1}\right) }},
\end{equation}

\noindent where $\upsilon $=$E\pi w^{2}/(2P_{g}\tau _{p})$ ($P_{g}$ is the
generation power, $E$ is the generation energy, $\omega _{p}$ is the pump
frequency, $\tau _{p}$ is the pulse width, $T_{cav}$ is the cavity period, $%
\sigma _{a}$ is the absorption cross-section, $I_{s}$ is the gain
saturation energy), $w$ is the generation mode beam size.
$\upsilon =\sqrt{\pi \diagup 2}$ for the pulse with the Gaussian
time-profile, 2 for the $sech$-shaped pulse and 1 for the CW (in
the latter case $\tau _{p}$=$T_{cav}$). The approximated solution
of Eq. (3) is based on the so-called aberration-less
approximation: the propagating field has the invariable
spatial-time profile, which is described by the set of the
$z$-dependent parameters. In the non-dissipative case this
approximation allows the variational approach providing rigorous
description of the Gaussian beam propagation outside the
parabolical approximation ~\cite{Anderson1}.

In the dissipative case we use the momentum method ~\cite{Vlasov} and
consider the momentums resulting from the symmetries of Eq. (3). The $%
a\rightarrow a\exp \left( {i\phi }\right) $ invariance, the transverse and
time translating invariance suggest the following momentums ~\cite{Akhmediev}%
:

\begin{gather}
T_{m,n}=\iint\limits_{\infty }{r^{m}t^{n}\left\vert a\right\vert ^{2}drdt,}%
\hfill  \\
J_{m.n}=\iint\limits_{\infty }{r^{m}t^{n}\left( {a\frac{{\partial a^{\ast }}%
}{{\partial t}}-a^{\ast }\frac{{\partial a}}{{\partial t}}}\right) drdt,}%
\hfill   \notag \\
M_{m.n}=\iint\limits_{\infty }{r^{m}t^{n}\left( {a\frac{{\partial a^{\ast }}%
}{{\partial r}}-a^{\ast }\frac{{\partial a}}{{\partial r}}}\right) drdt.}
\notag
\end{gather}

\noindent Like the variational approach we can substitute to Eqs. (3, 5) the
trial solution describing the ultrashort pulse. If we take the Gaussian
time-spatial profile $a\left( {z,r,t}\right) =W(r)\exp \left( {G(r)}{-\frac{{%
r^{2}}}{{2w(r)^{\prime }{}^{2}}}+ib(r)r^{2}-\frac{{t^{2}}}{{\tau (r)^{2}}}%
+i\psi (r)t^{2}}\right) $ ($W(r)$ is the complex amplitude, $2w^{\prime
}{}^{2}$=$w^{2}$, $G(r)$ is the pulse amplification parameter excepting the
geometrical magnification for the Gaussian beam), the equations describing
the evolution of the pulse and beam parameters are \cite{Kalash6}:

\begin{gather}
\frac{dw{^{\prime }}}{dz}=-\frac{2}{k}w^{\prime }\left( z\right) b\left(
z\right) -\frac{2\alpha w{^{\prime }}\left( z\right) {^{3}}}{w{%
_{p}^{2}\left( {1+\frac{{2w^{\prime }\left( z\right) ^{2}}}{{w_{p}^{2}}}}%
\right) ^{{3/2}}}}, \\
\frac{{d\tau \left( z\right) }}{{dz}}=\left[ {\frac{2}{{\tau \left( z\right)
}}-2\tau \left( z\right) ^{3}\psi \left( z\right) ^{2}}\right]
t_{f}^{2}+2\beta _{2}^{\prime }{}^{2}\tau \left( z\right) \psi \left(
z\right) ,  \notag \\
\frac{{dG\left( z\right) }}{{dz}}=\frac{\alpha }{{1+\frac{{2w^{\prime
}\left( z\right) ^{2}}}{{w_{p}^{2}}}}}-\frac{{2t_{f}^{2}}}{{\tau \left(
z\right) ^{2}}}-\beta _{2}^{\prime }{}^{2}\psi \left( z\right) ,  \notag \\
\frac{{db\left( z\right) }}{{dz}}=\frac{{2\vartheta }}{{w_{p}^{2}}}+\frac{{%
2b\left( z\right) ^{2}}}{k}+\frac{{\sqrt{2}P_{0}e^{2G\left( z\right) }}}{{%
\pi P_{cr}kw\left( z\right) ^{\prime }{}^{4}}}-\frac{1}{{2kw^{\prime }\left(
z\right) ^{4}}},  \notag \\
\frac{{d\psi \left( z\right) }}{{dz}}=2\beta _{2}^{\prime }{}^{2}\left( {%
\frac{1}{{\tau \left( z\right) ^{4}}}-\psi \left( z\right) ^{2}}\right) -%
\frac{{8t_{f}^{2}\psi \left( z\right) }}{{\tau \left( z\right) ^{2}}}+\frac{{%
2P_{0}e^{2G\left( z\right) }}}{{\pi P_{cr}w\left( z\right) ^{\prime
}{}^{2}\tau \left( z\right) ^{2}}},  \notag \\
P_{g}=P_{0}e^{2G\left( z\right) },  \notag
\end{gather}

\noindent where $P_{0}$ and $w_{0}^{\prime }$ are the power and the beam
size before the active medium, respectively. This system can be solved on
the basis of the fourth-order Runge-Kutta method (see Appendix 3; \emph{beta}%
, \emph{alpha}, \emph{gamma}, \emph{delta} and \emph{psi}
correspond to the right-hand side of Eqs. (6) for $b$, $w$, $G$,
$\tau $ and $\psi $, respectively).

\section{Conclusion}

\noindent Above considered models allow the different
generalizations. For example, in the framework of the
Ginzburg-Landau model the stimulated Raman scattering inside the
active medium can be taken into account immediately (by analogy
with the fiber optics). The high-order nonlinear Schr\"{o}dinger
equation can be generalized in order to take into account the
birefringence. The momentum method requires an additional analysis
for the dissipative propagation. Nevertheless, in the presented
form it can be used for the Kerr-lens mode-locked lasers
optimization. Outside the aberration-less approximation for the
field time-profile, the model allows most adequate description of
the mode-locked lasers, but the computational time can be enormous
in this case (for up-to-date desk-top computers).

The considered numerical codes were prepared as a result of the preliminary
analysis in the framework of the computer algebra system MAPLE (see
http://www.geocities.com/optomaplev).

\section{Acknowledgments}

This work was supported by Austrian National Science Fund under Project M688.

\bigskip

\section{Appendix 1:}

\noindent\ \ \ \ \ \
cccccccccccccccccccccccccccccccccccccccccccccccccccccccccccccccccccccccccc

c ULTRASHORT PULSE STABILITY

c IN THE KERR-LENS MODE-LOCKED LASER:

c ANALYSIS ON THE BASIS OF THE COMPLEX

c GINZBURG-LANDAU EQUATION

c

c V.L.Kalashnikov

c Photonics Institute, Technical University of Vienna

c e-mail: kalashnikov@tuwien.ac.at\qquad \qquad \qquad \qquad

c web-site: http://www.geocities.com/optomaplev\qquad \qquad \qquad \qquad
\qquad

cccccccccccccccccccccccccccccccccccccccccccccccccccccccccccccccccccccccccc

c

c\qquad PARAMETERS:

c

c\qquad \emph{alpha} is the saturated net-gain;

c\qquad \emph{beta} is the group-delay dispersion;

c\qquad \emph{D3} is the third-order dispersion;

c\qquad \emph{gamma} is the fast absorber modulation depth;

c\qquad \emph{delta} is the fast absorber saturation parameter.

c\qquad \emph{Vr} and \emph{Vi} are the real and imaginary parts of the
field, respectively;

c\qquad \emph{V} is the field intensity; \emph{VV} is the spectral intensity;

c\qquad \emph{Vmax} is the maximum pulse intensity;

c\qquad \emph{En} is the generation energy;

c\qquad \emph{width} is the pulse width;

c\qquad \emph{shift} is the spectrum maximum shift.

c

c\qquad The number of the considered steady-state pulses is defined by

c\qquad the number counter

c

c\qquad All values are dimensionless: the intensity is normalized to the
self-

c\qquad phase modulation coefficient, the time is normalized to the inverse

c\qquad bandwidth of the spectral filter, the propagation distance is
normalized

c\qquad to the cavity length

\qquad

\qquad REAL*8 Vr(4096),Vi(4096),V(4096),VV(4096),alpha,beta,gamma,delta

\qquad common /comin/ts(4096),ntab

\qquad

\qquad DOUBLE PRECISION rho,rho1,rho2,tau,C,S,X,Y,Argum,Fout,Omega,Vm

\qquad DATA Nt,Nst,Pi/4096,12,3.14159265358979323846264338d0/

\qquad

\qquad \qquad

\qquad

\qquad \qquad \qquad \qquad \qquad \qquad \qquad epsilon = 1.e-4

\qquad open(1,file='Landau\_G.dat')

\qquad write(1,*)'gamma, alpha, beta, D3, delta, Vmax, En, width,
shift'\qquad

\qquad \qquad \qquad \qquad \qquad gamma = 0.05

\qquad \qquad \qquad \qquad \qquad D3 = -150.

\qquad \qquad \qquad \qquad \qquad \qquad \qquad do j2=1,10

\qquad \qquad \qquad \qquad \qquad \qquad \qquad
alpha=(gamma-epsilon)*j2/10d0

\qquad \qquad \qquad \qquad \qquad \qquad \qquad do j3=1,100

\qquad \qquad \qquad \qquad \qquad \qquad \qquad beta=-j3

\qquad \qquad \qquad \qquad \qquad \qquad \qquad do j4=1,50

\qquad \qquad \qquad \qquad \qquad \qquad \qquad delta=10**(-2d0+4d0*j4/50d0)

\qquad Fout = 1d0/Nt

\qquad

c\qquad Initialization of the initial field

\qquad \qquad \qquad \qquad \qquad \qquad \qquad K = 0\qquad \qquad

\qquad tau=sqrt(gamma-alpha)

\qquad rho1 = sqrt(2d0)*tau/sqrt(gamma*delta)\ \ !from cubic Landau-Ginzburg

\qquad rho2 = sqrt(-beta)*tau\qquad \qquad \qquad \qquad ! from Schr\={o}%
dinger

\qquad if(rho1.gt.rho2)then

\qquad rho=rho1

\qquad else

\qquad rho=rho2

\qquad end if

\qquad \qquad

\qquad \qquad do j=1,Nt

\qquad \qquad \qquad Vr(j)=rho/cosh((j-2048)*tau)

\qquad \qquad \qquad \qquad Vi(j)=0.

\qquad \qquad \qquad end do

1\qquad \qquad \qquad \qquad \qquad \qquad \qquad K = K+1

c \ \ \ \ DISSIPATIVE PART

\qquad \qquad \qquad \qquad

c\qquad First amplification step

\qquad do I=1,Nt

\qquad

\qquad Vr(i) = Vr(i)*exp(0.5d0*alpha)

\qquad Vi(i) = Vi(i)*exp(0.5d0*alpha)

\qquad end do

c \ First filter action

\qquad call fftinn(Vr,Vi,Nt,Nst,1.) ! from Time to Frequency

\qquad \qquad do I=1,Nt

\qquad \qquad IF(I.LE.Nt/2+1)Is=I-1

\qquad \qquad IF(I.GE.Nt/2+2)Is=I-1-Nt

\qquad \qquad Omega=2.*Pi*Is*Fout

\qquad \qquad Argum=Omega**2 ! gain profile and action

\qquad \qquad Vr(I)=Vr(I)*exp(-0.5d0*Argum)

\qquad \qquad Vi(I)=Vi(I)*exp(-0.5d0*Argum)

\qquad \qquad end do

\qquad call fftinn(Vr,Vi,Nt,Nst,-1.) ! from Frequency to Time

c\qquad First nonlinear part's action

\qquad do I=1,Nt

\qquad \qquad Argum=gamma/(1d0 + delta*(Vr(I)**2+Vi(I)**2))

\qquad \qquad \qquad Vr(I)=Vr(I)*exp(-0.5d0*Argum)

\qquad \qquad \qquad Vi(I)=Vi(I)*exp(-0.5d0*Argum)

\qquad end do

c\qquad Second amplification step

\qquad do I=1,Nt

\qquad

\qquad Vr(i) = Vr(i)*exp(0.5d0*alpha)

\qquad Vi(i) = Vi(i)*exp(0.5d0*alpha)

\qquad end do

c \qquad Second filter action

\qquad call fftinn(Vr,Vi,Nt,Nst,1.) ! from Time to Frequency

\qquad \qquad do I=1,Nt

\qquad \qquad IF(I.LE.Nt/2+1)Is=I-1

\qquad \qquad IF(I.GE.Nt/2+2)Is=I-1-Nt

\qquad \qquad Omega=2.*Pi*Is*Fout

\qquad \qquad Argum=Omega**2 ! gain profile and action

\qquad \qquad Vr(I)=Vr(I)*exp(-0.5d0*Argum)

\qquad \qquad Vi(I)=Vi(I)*exp(-0.5d0*Argum)

\qquad \qquad end do

\qquad call fftinn(Vr,Vi,Nt,Nst,-1.) ! from Frequency to Time

c\qquad Second nonlinear part's action

\qquad do I=1,Nt

\qquad \qquad Argum=gamma/(1d0 + delta*(Vr(I)**2+Vi(I)**2))

\qquad \qquad \qquad Vr(I)=Vr(I)*exp(-0.5d0*Argum)

\qquad \qquad \qquad Vi(I)=Vi(I)*exp(-0.5d0*Argum)

\qquad end do

c \ HAMILTONIAN PART

c\qquad First dispersion step

\qquad call fftinn(Vr,Vi,Nt,Nst,1.) ! from Time to Frequency

\qquad \qquad do I=1,Nt

\qquad \qquad IF(I.LE.Nt/2+1)Is=I-1

\qquad \qquad IF(I.GE.Nt/2+2)Is=I-1-Nt

\qquad \qquad Omega=2.*Pi*Is*Fout

\qquad \qquad C=cos(0.5d0*((beta/2d0)*Omega**2 - (D3/6d0)*Omega**3))

\qquad \qquad S=sin(0.5d0*((beta/2d0)*Omega**2 - (D3/6d0)*Omega**3))

\qquad \qquad X=Vr(I)

\qquad \qquad Y=Vi(I)

\qquad \qquad Vr(I)=X*C+Y*S

\qquad \qquad Vi(I)=Y*C-X*S

\qquad \qquad end do

\qquad call fftinn(Vr,Vi,Nt,Nst,-1.) ! from Frequency to Time

c\qquad First self-phase modulation step

\qquad do I=1,Nt

\qquad \qquad Argum=0.5d0*(Vr(I)**2+Vi(I)**2)

\qquad \qquad \qquad C=cos(Argum)

\qquad \qquad \qquad \qquad S=sin(Argum)

\qquad \qquad \qquad \qquad \qquad X=Vr(I)

\qquad \qquad \qquad \qquad Y=Vi(I)

\qquad \qquad \qquad Vr(I)=X*C+Y*S

\qquad \qquad Vi(I)=Y*C-X*S

\qquad end do

c\qquad Second dispersion step

\qquad call fftinn(Vr,Vi,Nt,Nst,1.) ! from Time to Frequency

\qquad \qquad do I=1,Nt

\qquad \qquad IF(I.LE.Nt/2+1)Is=I-1

\qquad \qquad IF(I.GE.Nt/2+2)Is=I-1-Nt

\qquad \qquad Omega=2.*Pi*Is*Fout

\qquad \qquad C=cos(0.5d0*((beta/2d0)*Omega**2 - (D3/6d0)*Omega**3))

\qquad \qquad S=sin(0.5d0*((beta/2d0)*Omega**2 - (D3/6d0)*Omega**3))

\qquad \qquad X=Vr(I)

\qquad \qquad Y=Vi(I)

\qquad \qquad Vr(I)=X*C+Y*S

\qquad \qquad Vi(I)=Y*C-X*S

\qquad \qquad end do

\qquad call fftinn(Vr,Vi,Nt,Nst,-1.) ! from Frequency to Time

c\qquad Second self-phase modulation step

\qquad do I=1,Nt

\qquad \qquad Argum=0.5d0*(Vr(I)**2+Vi(I)**2)

\qquad \qquad \qquad C=cos(Argum)

\qquad \qquad \qquad \qquad S=sin(Argum)

\qquad \qquad \qquad \qquad \qquad X=Vr(I)

\qquad \qquad \qquad \qquad Y=Vi(I)

\qquad \qquad \qquad Vr(I)=X*C+Y*S

\qquad \qquad Vi(I)=Y*C-X*S

\qquad end do

c Analysis of Generation Field \qquad

\qquad call fftinn(Vr,Vi,Nt,Nst,1.) ! from Time to Frequency

\qquad do i=1,nt

\qquad if(i.le.nt/2)then

\qquad j=nt/2+i

\qquad VV(j)=Vr(i)**2 + Vi(i)**2

\qquad else

\qquad j=i-nt/2

\qquad VV(j)=Vr(i)**2 + Vi(i)**2

\qquad end if

\qquad end do

\qquad call fftinn(Vr,Vi,Nt,Nst,-1.) ! from Frequency to Time

\qquad Sm=VV(1)

\qquad do 6 I=1,NT

\qquad if(VV(I).LT.Sm)goto 6

\qquad Sm=VV(I)

\qquad Ism=I

6\qquad continue

\qquad shift=Ism-Nt/2

\qquad do I=1,NT

\qquad V(I)=Vr(I)**2+Vi(I)**2

\qquad end do

\qquad

\qquad

\qquad Vm=V(1)

\qquad do 2 I=1,NT

\qquad if(V(I).LT.Vm)goto 2

\qquad Vm=V(I)

\qquad Im=I

2\qquad continue

\qquad if(k.eq.9000)cont\_p=Vm\qquad

\qquad if(Vm.lt.1e-10)goto 5

\qquad if(k.lt.10000)goto 1

c Analysis of Generation Field \qquad

\qquad do I=1,NT

\qquad IF(V(I).ge.Vm/2.and.V(I-1).le.Vm/2)h1=(2.*I-1)/2.

\qquad IF(V(I).le.Vm/2.and.V(I-1).ge.Vm/2)h2=(2.*I-1)/2.

\qquad end do

\qquad \qquad En = 0.

\qquad \qquad do I=1,Nt

\qquad \qquad En=En+V(i)

\qquad \qquad end do

\qquad \qquad \qquad \qquad if(Vm.gt.0.)then

\qquad \qquad \qquad \qquad stab=abs(cont\_p-Vm)/Vm

\qquad \qquad \qquad \qquad else

\qquad \qquad \qquad \qquad stab=1.

\qquad \qquad \qquad \qquad end if

\qquad \qquad \qquad Vmax=Vm

c PULSE NUMBER COUNTER

call fftinn(Vr,Vi,Nt,Nst,1.) ! from Time to Frequency

\qquad

c Filter

\qquad do I=1,Nt

\qquad \qquad IF(I.LE.Nt/2+1)Is=I-1

\qquad \qquad IF(I.GE.Nt/2+2)Is=I-1-Nt

\qquad \qquad Omega=2.*Pi*Is*Fout

\qquad \qquad Argum=(0.1*(h2-h1))**2*Omega**2

\qquad \qquad Vr(I)=Vr(I)*exp(-Argum)

\qquad \qquad Vi(I)=Vi(I)*exp(-Argum)

\qquad end do

\qquad

\qquad

call fftinn(Vr,Vi,Nt,Nst,-1.) ! from Frequency to Time\qquad \qquad

\qquad \qquad

\qquad \qquad \qquad \qquad

\qquad do I=1,NT

\qquad V(I)=Vr(I)**2+Vi(I)**2

\qquad end do

\qquad Vm=V(1)

\qquad do 3 I=1,NT

\qquad if(V(I).LT.Vm)goto 3

\qquad Vm=V(I)

\qquad Im=I

3\qquad continue

\qquad

\qquad nm=0

\qquad do i=2,nt-1

\qquad if(V(i).ge.V(i-1).and.V(i).gt.Vm/10.)then

\qquad if(V(i+1).lt.V(i))nm=nm+1

\qquad else

\qquad end if

\qquad end do

\qquad \qquad width=(h2-h1)

\qquad if(nm.eq.1.and.stab.lt.0.01)then

\qquad write(1,fmt=4)gamma,alpha,beta,D3,delta,Vmax,En,width,shift\qquad

4\qquad format(f4.2,1x,f5.3,1x,f6.1,1x,f6.1,1x,f5.1,1x,f7.4,1x,f7.3,1x,

\#f7.1,1x,f7.1)

\qquad else

\qquad end if

5\qquad \qquad \qquad \qquad \qquad \qquad \qquad end do\qquad

\qquad \qquad \qquad \qquad \qquad \qquad \qquad end do

\qquad \qquad \qquad \qquad \qquad \qquad \qquad end do

\qquad close(1)

\qquad end

\qquad \qquad \qquad \qquad \qquad \qquad

c\qquad Fast Fourier transformation

\qquad SUBROUTINE FFTINN(XB,XM,NT,NST,SIGN)

\qquad REAL*8 XB(NT),XM(NT)

\qquad COMMON/COMIN/TS(4096),NTAB

\qquad DOUBLE PRECISION XB1,XB2,XM1,XM2,PRIR,W1,W2,FP,TS

\qquad IF(NTAB.GT.0)GOTO 30

\qquad NTAB=4096

\qquad PRIR=3.14159265358979323846264338d0/NTAB

\qquad DO 20 I=1,NTAB

\qquad FP=PRIR*(I-1)

20 \ TS(I)=SIN(FP)

30 \ CONTINUE

\qquad LEC1=NT-1

\qquad LEC2=NT/2

\qquad J=1

\qquad DO 10 I=1,LEC1

\qquad IF(I.GE.J)GO TO 8

\qquad XB1=XB(J)

\qquad XM1=XM(J)

\qquad XB(J)=XB(I)

\qquad XM(J)=XM(I) \qquad

\qquad XB(I)=XB1 \qquad

\qquad XM(I)=XM1

8 \ \ \ \ L=LEC2

9 \ \ \ \ IF(L.GE.J)GOTO 10

\qquad J=J-L

\qquad L=L/2

\qquad GO TO 9

10 \ \ J=J+L

\qquad INCP=NTAB*2

\qquad NSDV=NTAB/2

\qquad JLI=1

\qquad JKI=1

\qquad KLI=NT

\qquad DO 3 I=1,NST

\qquad JKI=JKI+JKI

\qquad KLI=KLI/2

\qquad INCP=INCP/2

\qquad INI=1

\qquad DO 2 J=1,JLI

\qquad LEC1=J

\qquad W2=TS(INI)

\qquad IF(INI-NSDV)5,5,6

5 \ \ \ W1=TS(INI+NSDV)

\qquad GO TO 7

6 \ \ \ \ W1=-TS(INI-NSDV)

7 \ \ \ IF(SIGN.GT.0.)W2=-W2

\qquad DO 1 K=1,KLI

\qquad LEC2=LEC1+JLI

\qquad XB1=XB(LEC1)

\qquad XM1=XM(LEC1)

\qquad XB2=W1*XB(LEC2)-W2*XM(LEC2)

\qquad XM2=W1*XM(LEC2)+W2*XB(LEC2)

\qquad XB(LEC1)=XB1+XB2

\qquad XM(LEC1)=XM1+XM2

\qquad XB(LEC2)=XB1-XB2

\qquad XM(LEC2)=XM1-XM2

1 \ \ \ \ LEC1=LEC1+JKI

2 \ \ \ \ INI=INI+INCP

3 \ \ \ \ JLI=JLI+JLI

\qquad IF(SIGN.LT.0.)RETURN

\qquad FP=1./NT

\qquad DO 4 I=1,NT

\qquad XB(I)=XB(I)*FP

4 \ \ \ \ XM(I)=XM(I)*FP

\qquad RETURN

\qquad END

\section{\qquad Appendix 2:}

\qquad \noindent
cccccccccccccccccccccccccccccccccccccccccccccccccccccccccccccccccccccccccc

cSPECTRAL CONTINUUM GENERATION IN THE TAPERED FIBER:

c\qquad \qquad GENERALIZED SCHROEDINGER EQUATION

c

c\qquad \qquad \qquad \qquad \qquad \qquad V.L.Kalashnikov

c\qquad \qquad Photonics Institute, Technical University of Vienna

c\qquad \qquad \qquad \qquad \qquad e-mail: kalashnikov@tuwien.ac.at

c\qquad \qquad \qquad web-site: http://www.geocities.com/optomaplev

cccccccccccccccccccccccccccccccccccccccccccccccccccccccccccccccccccccccccc

c

c\qquad The description of the program can be found on
http://www.geocities.com/

c\qquad optomaplev/programs/fortran.html

c

c\qquad PARAMETERS:

c

c Amp is the initial pulse intensity (1 for the tapered section)

c ar\_begin is defined by the ration of the tapered fiber and single-mode

c fiber cross-sections

c width is the initial pulse width (in fs)

c D\_begin and D3\_begin are the group-velocity and third-order dispersions

c of the single-mode fiber

c D\_end and D3\_end are the group-velocity and third-order dispersions

c of the tapered fiber

c All dispersions are defined through the dispersion lengths normalized to

c the nonlinear length in the tapered fiber

c N\_steps and N\_t are 1000*L/Lnl, where L is the length of the transient

c fiber sector or the tapered sector, respectively, Lnl is the

c nonlinear length of the single-mode or tipered fiber, respectively

\qquad

\qquad REAL*8 Vr(8192),Vi(8192),V(8192),R(8192)

\qquad REAL*8 Vr\_out(8192),Vi\_out(8192)

\qquad INTEGER N\_steps,N\_t

\qquad common /comin/ts(8192),ntab

\qquad

\qquad DOUBLE PRECISION Amp,width,C,S,X,Y,Argum,Fout,Omega,Vm,n2eff

\qquad DATA Nt,Nst,Pi,Dx/8192,13,3.14159265358979323846264338d0,1d-3/

\qquad DATA Fr,T1,T2/0.15d0,12.2d0,32d0/

\qquad

\qquad \qquad

\qquad

\qquad open(1,file='phase.dat')

\qquad open(2,file='spectrum.dat')

\qquad open(3,file='intensity.dat')

\qquad Fout = 1d0/Nt\qquad

c\qquad Raman responce function

\qquad do i=1,Nt

\qquad \qquad R(i) = ((T1**2+T2**2)/(T1*T2**2))*exp(-(i-1)/T2)*sin((i-1)/T1)

\qquad end do

c\qquad Initialization of the initial field

\qquad \qquad \qquad \qquad \qquad \qquad \qquad K = 0\qquad \qquad

\qquad width=28.36d0

\qquad

\qquad

\qquad D\_begin = -width**2/7.55d0

\qquad D\_end = width**2/0.53d0

\qquad D3\_begin = -0.58d0*D\_begin*width

\qquad D3\_end = -0.78d0*D\_end*width

\qquad ar\_begin = 0.0149

\qquad N\_steps = 3774

\qquad \qquad N\_t = 16992

\qquad \qquad \qquad \qquad \qquad \qquad AAA =
ar\_begin**(-1./(2.*N\_steps))

\qquad D = D\_begin

\qquad D3 = D3\_begin

\qquad \qquad \qquad \qquad \qquad \qquad

\qquad Amp = sqrt(ar\_begin)

\qquad \qquad

\qquad \qquad do j=1,Nt

\qquad \qquad \qquad Vr(j)=Amp/cosh((j-4096)/width)

\qquad \qquad \qquad \qquad Vi(j)=0.

\qquad \qquad \qquad end do

\qquad do j=1,Nt

\qquad write(3,*)k,j,Vr(j)**2+Vi(j)**2

\qquad end do

call fftinn(Vr,Vi,Nt,Nst,1.) ! from Time to Frequency

\qquad do i=1,nt

\qquad if(i.le.nt/2)then

\qquad j=nt/2+i

\qquad V(j)=Vr(i)**2 + Vi(i)**2

\qquad Vr\_out(j)=Vr(i)

\qquad Vi\_out(j)=Vi(i)

\qquad else

\qquad j=i-nt/2

\qquad V(j)=Vr(i)**2 + Vi(i)**2

\qquad Vr\_out(j)=Vr(i)

\qquad Vi\_out(j)=Vi(i)

\qquad end if

\qquad end do

\qquad call fftinn(Vr,Vi,Nt,Nst,-1.) ! from Frequency to Time

\qquad do i=1,Nt

\qquad write(1,*)k,i,Vr\_out(i),Vi\_out(i)

\qquad end do

\qquad do j=1,Nt

\qquad write(2,*)k,j,V(j)

\qquad end do

c\qquad FIRST TRANSIENT SECTOR

1\qquad \qquad \qquad \qquad \qquad \qquad \qquad K = K+1

\qquad write(*,*)K

\qquad \qquad \qquad \qquad

c\qquad First dispersion step

call fftinn(Vr,Vi,Nt,Nst,1.) ! from Time to Frequency

\qquad do I=1,Nt

\qquad \qquad IF(I.LE.Nt/2+1)Is=I-1

\qquad \qquad IF(I.GE.Nt/2+2)Is=I-1-Nt

\qquad \qquad Omega=2.*Pi*Is*Fout

\qquad \qquad C=cos(Dx*0.5d0*((D/2d0)*Omega**2 - (D3/6d0)*Omega**3))

\qquad \qquad S=sin(Dx*0.5d0*((D/2d0)*Omega**2 - (D3/6d0)*Omega**3))

\qquad \qquad X=Vr(I)

\qquad \qquad Y=Vi(I)

\qquad \qquad Vr(I)= X*C - Y*S

\qquad \qquad Vi(I)= Y*C + X*S

\qquad end do

call fftinn(Vr,Vi,Nt,Nst,-1.) ! from Frequency to Time

c\qquad First self-phase modulation step

\qquad do I=1,Nt

\qquad \qquad Argum=Dx*0.5d0*(1d0-Fr)*(Vr(I)**2+Vi(I)**2)

\qquad \qquad \qquad C=cos(Argum)

\qquad \qquad \qquad \qquad S=sin(Argum)

\qquad \qquad \qquad \qquad \qquad X=Vr(I)

\qquad \qquad \qquad \qquad Y=Vi(I)

\qquad \qquad \qquad Vr(I)=X*C - Y*S

\qquad \qquad Vi(I)=Y*C + X*S

\qquad end do

c\qquad First Raman step

\qquad do I=1,Nt

\qquad \qquad

\qquad Argum = 0d0

\qquad \qquad do J=1,I

\qquad \qquad Argum=Argum + R(I-J+1)*(Vr(J)**2+Vi(J)**2)

\qquad \qquad end do

\qquad \qquad \qquad C=cos(Dx*0.5d0*Fr*Argum)

\qquad \qquad \qquad \qquad S=sin(Dx*0.5d0*Fr*Argum)

\qquad \qquad \qquad \qquad \qquad X=Vr(I)

\qquad \qquad \qquad \qquad Y=Vi(I)

\qquad \qquad \qquad Vr(I)=X*C - Y*S

\qquad \qquad Vi(I)=Y*C + X*S

\qquad \qquad

\qquad end do

c\qquad Second dispersion step

call fftinn(Vr,Vi,Nt,Nst,1.) ! from Time to Frequency

\qquad do I=1,Nt

\qquad \qquad IF(I.LE.Nt/2+1)Is=I-1

\qquad \qquad IF(I.GE.Nt/2+2)Is=I-1-Nt

\qquad \qquad Omega=2.*Pi*Is*Fout

\qquad \qquad C=cos(Dx*0.5d0*((D/2d0)*Omega**2 - (D3/6d0)*Omega**3))

\qquad \qquad S=sin(Dx*0.5d0*((D/2d0)*Omega**2 - (D3/6d0)*Omega**3))

\qquad \qquad X=Vr(I)

\qquad \qquad Y=Vi(I)

\qquad \qquad Vr(I)= X*C - Y*S

\qquad \qquad Vi(I)= Y*C + X*S

\qquad end do

call fftinn(Vr,Vi,Nt,Nst,-1.) ! from Frequency to Time

c\qquad Second self-phase modulation step

\qquad do I=1,Nt

\qquad \qquad Argum=Dx*0.5d0*(1d0-Fr)*(Vr(I)**2+Vi(I)**2)

\qquad \qquad \qquad C=cos(Argum)

\qquad \qquad \qquad \qquad S=sin(Argum)

\qquad \qquad \qquad \qquad \qquad X=Vr(I)

\qquad \qquad \qquad \qquad Y=Vi(I)

\qquad \qquad \qquad Vr(I)=X*C - Y*S

\qquad \qquad Vi(I)=Y*C + X*S

\qquad end do

c\qquad Second Raman step

\qquad do I=1,Nt

\qquad \qquad

\qquad Argum = 0d0

\qquad \qquad do J=1,I

\qquad \qquad Argum=Argum + R(I-J+1)*(Vr(J)**2+Vi(J)**2)

\qquad \qquad end do

\qquad \qquad \qquad C=cos(Dx*0.5d0*Fr*Argum)

\qquad \qquad \qquad \qquad S=sin(Dx*0.5d0*Fr*Argum)

\qquad \qquad \qquad \qquad \qquad X=Vr(I)

\qquad \qquad \qquad \qquad Y=Vi(I)

\qquad \qquad \qquad Vr(I)=X*C - Y*S

\qquad \qquad Vi(I)=Y*C + X*S

\qquad end do

\qquad do I=1,Nt

\qquad Vr(i) = Vr(i)*AAA

\qquad Vi(i) = Vi(i)*AAA

\qquad end do

\qquad D = D\_begin + K*(D\_end - D\_begin)/N\_steps

\qquad D3 = D3\_begin + K*(D3\_end - D3\_begin)/N\_steps

\qquad if((K/1000)*1000.eq.K)then

\qquad do j=1,Nt

\qquad write(3,*)k,j,Vr(j)**2+Vi(j)**2

\qquad end do

call fftinn(Vr,Vi,Nt,Nst,1.) ! from Time to Frequency

\qquad do i=1,nt

\qquad if(i.le.nt/2)then

\qquad j=nt/2+i

\qquad V(j)=Vr(i)**2 + Vi(i)**2

\qquad Vr\_out(j)=Vr(i)

\qquad Vi\_out(j)=Vi(i)

\qquad else

\qquad j=i-nt/2

\qquad V(j)=Vr(i)**2 + Vi(i)**2

\qquad Vr\_out(j)=Vr(i)

\qquad Vi\_out(j)=Vi(i)

\qquad end if

\qquad end do

\qquad call fftinn(Vr,Vi,Nt,Nst,-1.) ! from Frequency to Time

\qquad do i=1,Nt

\qquad write(1,*)k,i,Vr\_out(i),Vi\_out(i)

\qquad end do

\qquad do j=1,Nt

\qquad write(2,*)k,j,V(j)

\qquad end do

\qquad else

\qquad end if

\qquad if(k.lt.N\_steps)goto 1

c\qquad TAPERED SECTOR

\qquad \qquad \qquad \qquad \qquad \qquad \qquad K = K+1

\qquad write(*,*)K

\qquad \qquad D = D\_end

\qquad \qquad D3 = D3\_end

\qquad \qquad \qquad \qquad

c\qquad First dispersion step

call fftinn(Vr,Vi,Nt,Nst,1.) ! from Time to Frequency

\qquad do I=1,Nt

\qquad \qquad IF(I.LE.Nt/2+1)Is=I-1

\qquad \qquad IF(I.GE.Nt/2+2)Is=I-1-Nt

\qquad \qquad Omega=2.*Pi*Is*Fout

\qquad \qquad C=cos(Dx*0.5d0*((D/2d0)*Omega**2 - (D3/6d0)*Omega**3))

\qquad \qquad S=sin(Dx*0.5d0*((D/2d0)*Omega**2 - (D3/6d0)*Omega**3))

\qquad \qquad X=Vr(I)

\qquad \qquad Y=Vi(I)

\qquad \qquad Vr(I)= X*C - Y*S

\qquad \qquad Vi(I)= Y*C + X*S

\qquad end do

call fftinn(Vr,Vi,Nt,Nst,-1.) ! from Frequency to Time

c\qquad First self-phase modulation step

\qquad do I=1,Nt

\qquad \qquad Argum=Dx*0.5d0*(1d0-Fr)*(Vr(I)**2+Vi(I)**2)

\qquad \qquad \qquad C=cos(Argum)

\qquad \qquad \qquad \qquad S=sin(Argum)

\qquad \qquad \qquad \qquad \qquad X=Vr(I)

\qquad \qquad \qquad \qquad Y=Vi(I)

\qquad \qquad \qquad Vr(I)=X*C - Y*S

\qquad \qquad Vi(I)=Y*C + X*S

\qquad end do

c\qquad First Raman step

\qquad do I=1,Nt

\qquad \qquad

\qquad Argum = 0d0

\qquad \qquad do J=1,I

\qquad \qquad Argum=Argum + R(I-J+1)*(Vr(J)**2+Vi(J)**2)

\qquad \qquad end do

\qquad \qquad \qquad C=cos(Dx*0.5d0*Fr*Argum)

\qquad \qquad \qquad \qquad S=sin(Dx*0.5d0*Fr*Argum)

\qquad \qquad \qquad \qquad \qquad X=Vr(I)

\qquad \qquad \qquad \qquad Y=Vi(I)

\qquad \qquad \qquad Vr(I)=X*C - Y*S

\qquad \qquad Vi(I)=Y*C + X*S

\qquad \qquad

\qquad end do

c\qquad Second dispersion step

call fftinn(Vr,Vi,Nt,Nst,1.) ! from Time to Frequency

\qquad do I=1,Nt

\qquad \qquad IF(I.LE.Nt/2+1)Is=I-1

\qquad \qquad IF(I.GE.Nt/2+2)Is=I-1-Nt

\qquad \qquad Omega=2.*Pi*Is*Fout

\qquad \qquad C=cos(Dx*0.5d0*((D/2d0)*Omega**2 - (D3/6d0)*Omega**3))

\qquad \qquad S=sin(Dx*0.5d0*((D/2d0)*Omega**2 - (D3/6d0)*Omega**3))

\qquad \qquad X=Vr(I)

\qquad \qquad Y=Vi(I)

\qquad \qquad Vr(I)= X*C - Y*S

\qquad \qquad Vi(I)= Y*C + X*S

\qquad end do

call fftinn(Vr,Vi,Nt,Nst,-1.) ! from Frequency to Time

c\qquad Second self-phase modulation step

\qquad do I=1,Nt

\qquad \qquad Argum=Dx*0.5d0*(1d0-Fr)*(Vr(I)**2+Vi(I)**2)

\qquad \qquad \qquad C=cos(Argum)

\qquad \qquad \qquad \qquad S=sin(Argum)

\qquad \qquad \qquad \qquad \qquad X=Vr(I)

\qquad \qquad \qquad \qquad Y=Vi(I)

\qquad \qquad \qquad Vr(I)=X*C - Y*S

\qquad \qquad Vi(I)=Y*C + X*S

\qquad end do

c\qquad Second Raman step

\qquad do I=1,Nt

\qquad \qquad

\qquad Argum = 0d0

\qquad \qquad do J=1,I

\qquad \qquad Argum=Argum + R(I-J+1)*(Vr(J)**2+Vi(J)**2)

\qquad \qquad end do

\qquad \qquad \qquad C=cos(Dx*0.5d0*Fr*Argum)

\qquad \qquad \qquad \qquad S=sin(Dx*0.5d0*Fr*Argum)

\qquad \qquad \qquad \qquad \qquad X=Vr(I)

\qquad \qquad \qquad \qquad Y=Vi(I)

\qquad \qquad \qquad Vr(I)=X*C - Y*S

\qquad \qquad Vi(I)=Y*C + X*S

\qquad end do

\qquad if((K/1000)*1000.eq.K)then

\qquad do j=1,Nt

\qquad write(3,*)k,j,Vr(j)**2+Vi(j)**2

\qquad end do

call fftinn(Vr,Vi,Nt,Nst,1.) ! from Time to Frequency

\qquad do i=1,nt

\qquad if(i.le.nt/2)then

\qquad j=nt/2+i

\qquad V(j)=Vr(i)**2 + Vi(i)**2

\qquad Vr\_out(j)=Vr(i)

\qquad Vi\_out(j)=Vi(i)

\qquad else

\qquad j=i-nt/2

\qquad V(j)=Vr(i)**2 + Vi(i)**2

\qquad Vr\_out(j)=Vr(i)

\qquad Vi\_out(j)=Vi(i)

\qquad end if

\qquad end do

\qquad call fftinn(Vr,Vi,Nt,Nst,-1.) ! from Frequency to Time

\qquad do i=1,Nt

\qquad write(1,*)k,i,Vr\_out(i),Vi\_out(i)

\qquad end do

\qquad do j=1,Nt

\qquad write(2,*)k,j,V(j)

\qquad end do

\qquad else

\qquad end if

\qquad if(k.lt.N\_steps+N\_t)goto 2

c\qquad SECOND TRANSIENT SECTOR

\qquad \qquad \qquad \qquad \qquad \qquad \qquad K = K+1

\qquad write(*,*)K

\qquad \qquad \qquad \qquad

c\qquad First dispersion step

call fftinn(Vr,Vi,Nt,Nst,1.) ! from Time to Frequency

\qquad do I=1,Nt

\qquad \qquad IF(I.LE.Nt/2+1)Is=I-1

\qquad \qquad IF(I.GE.Nt/2+2)Is=I-1-Nt

\qquad \qquad Omega=2.*Pi*Is*Fout

\qquad \qquad C=cos(Dx*0.5d0*((D/2d0)*Omega**2 - (D3/6d0)*Omega**3))

\qquad \qquad S=sin(Dx*0.5d0*((D/2d0)*Omega**2 - (D3/6d0)*Omega**3))

\qquad \qquad X=Vr(I)

\qquad \qquad Y=Vi(I)

\qquad \qquad Vr(I)= X*C - Y*S

\qquad \qquad Vi(I)= Y*C + X*S

\qquad end do

call fftinn(Vr,Vi,Nt,Nst,-1.) ! from Frequency to Time

c\qquad First self-phase modulation step

\qquad do I=1,Nt

\qquad \qquad Argum=Dx*0.5d0*(1d0-Fr)*(Vr(I)**2+Vi(I)**2)

\qquad \qquad \qquad C=cos(Argum)

\qquad \qquad \qquad \qquad S=sin(Argum)

\qquad \qquad \qquad \qquad \qquad X=Vr(I)

\qquad \qquad \qquad \qquad Y=Vi(I)

\qquad \qquad \qquad Vr(I)=X*C - Y*S

\qquad \qquad Vi(I)=Y*C + X*S

\qquad end do

c\qquad First Raman step

\qquad do I=1,Nt

\qquad \qquad

\qquad Argum = 0d0

\qquad \qquad do J=1,I

\qquad \qquad Argum=Argum + R(I-J+1)*(Vr(J)**2+Vi(J)**2)

\qquad \qquad end do

\qquad \qquad \qquad C=cos(Dx*0.5d0*Fr*Argum)

\qquad \qquad \qquad \qquad S=sin(Dx*0.5d0*Fr*Argum)

\qquad \qquad \qquad \qquad \qquad X=Vr(I)

\qquad \qquad \qquad \qquad Y=Vi(I)

\qquad \qquad \qquad Vr(I)=X*C - Y*S

\qquad \qquad Vi(I)=Y*C + X*S

\qquad \qquad

\qquad end do

c\qquad Second dispersion step

call fftinn(Vr,Vi,Nt,Nst,1.) ! from Time to Frequency

\qquad do I=1,Nt

\qquad \qquad IF(I.LE.Nt/2+1)Is=I-1

\qquad \qquad IF(I.GE.Nt/2+2)Is=I-1-Nt

\qquad \qquad Omega=2.*Pi*Is*Fout

\qquad \qquad C=cos(Dx*0.5d0*((D/2d0)*Omega**2 - (D3/6d0)*Omega**3))

\qquad \qquad S=sin(Dx*0.5d0*((D/2d0)*Omega**2 - (D3/6d0)*Omega**3))

\qquad \qquad X=Vr(I)

\qquad \qquad Y=Vi(I)

\qquad \qquad Vr(I)= X*C - Y*S

\qquad \qquad Vi(I)= Y*C + X*S

\qquad end do

call fftinn(Vr,Vi,Nt,Nst,-1.) ! from Frequency to Time

c\qquad Second self-phase modulation step

\qquad do I=1,Nt

\qquad \qquad Argum=Dx*0.5d0*(1d0-Fr)*(Vr(I)**2+Vi(I)**2)

\qquad \qquad \qquad C=cos(Argum)

\qquad \qquad \qquad \qquad S=sin(Argum)

\qquad \qquad \qquad \qquad \qquad X=Vr(I)

\qquad \qquad \qquad \qquad Y=Vi(I)

\qquad \qquad \qquad Vr(I)=X*C - Y*S

\qquad \qquad Vi(I)=Y*C + X*S

\qquad end do

c\qquad Second Raman step

\qquad do I=1,Nt

\qquad \qquad

\qquad Argum = 0d0

\qquad \qquad do J=1,I

\qquad \qquad Argum=Argum + R(I-J+1)*(Vr(J)**2+Vi(J)**2)

\qquad \qquad end do

\qquad \qquad \qquad C=cos(Dx*0.5d0*Fr*Argum)

\qquad \qquad \qquad \qquad S=sin(Dx*0.5d0*Fr*Argum)

\qquad \qquad \qquad \qquad \qquad X=Vr(I)

\qquad \qquad \qquad \qquad Y=Vi(I)

\qquad \qquad \qquad Vr(I)=X*C - Y*S

\qquad \qquad Vi(I)=Y*C + X*S

\qquad end do

\qquad do I=1,Nt

\qquad Vr(i) = Vr(i)/AAA

\qquad Vi(i) = Vi(i)/AAA

\qquad end do

\qquad D = D\_end + (K-N\_steps-N\_t)*(D\_begin - D\_end)/N\_steps

\qquad D3 = D3\_end + (K-N\_steps-N\_t)*(D3\_begin - D3\_end)/N\_steps

\qquad if((K/1000)*1000.eq.K)then

\qquad do j=1,Nt

\qquad write(3,*)k,j,Vr(j)**2+Vi(j)**2

\qquad end do

call fftinn(Vr,Vi,Nt,Nst,1.) ! from Time to Frequency

\qquad do i=1,nt

\qquad if(i.le.nt/2)then

\qquad j=nt/2+i

\qquad V(j)=Vr(i)**2 + Vi(i)**2

\qquad Vr\_out(j)=Vr(i)

\qquad Vi\_out(j)=Vi(i)

\qquad else

\qquad j=i-nt/2

\qquad V(j)=Vr(i)**2 + Vi(i)**2

\qquad Vr\_out(j)=Vr(i)

\qquad Vi\_out(j)=Vi(i)

\qquad end if

\qquad end do

\qquad call fftinn(Vr,Vi,Nt,Nst,-1.) ! from Frequency to Time

\qquad do i=1,Nt

\qquad write(1,*)k,i,Vr\_out(i),Vi\_out(i)\qquad

\qquad end do

\qquad do j=1,Nt

\qquad write(2,*)k,j,V(j)\qquad \qquad \qquad \qquad \qquad

\qquad end do

\qquad else

\qquad end if

\qquad if(k.lt.N\_t + 2*N\_steps)goto 3

\qquad do j=1,Nt

\qquad write(3,*)k,j,Vr(j)**2+Vi(j)**2\qquad \qquad

\qquad end do

call fftinn(Vr,Vi,Nt,Nst,1.) ! from Time to Frequency

\qquad do i=1,nt

\qquad if(i.le.nt/2)then

\qquad j=nt/2+i

\qquad V(j)=Vr(i)**2 + Vi(i)**2

\qquad Vr\_out(j)=Vr(i)

\qquad Vi\_out(j)=Vi(i)

\qquad else

\qquad j=i-nt/2

\qquad V(j)=Vr(i)**2 + Vi(i)**2

\qquad Vr\_out(j)=Vr(i)

\qquad Vi\_out(j)=Vi(i)

\qquad end if

\qquad end do

\qquad call fftinn(Vr,Vi,Nt,Nst,-1.) ! from Frequency to Time

\qquad do i=1,Nt

\qquad write(1,*)k,i,Vr\_out(i),Vi\_out(i)

\qquad end do

\qquad do j=1,Nt

\qquad write(2,*)k,j,V(j)

\qquad end do

\qquad

\qquad close(1)

\qquad close(2)

\qquad close(3)

\qquad end

c\qquad Fast Fourier transformation

\qquad \emph{see Appendix 1}

\section{Appendix 3}

cccccccccccccccccccccccccccccccccccccccccccccccccccccccccccccccccccccccccc

c ULTRASHORT PULSE STABILITY IN THE KERR-LENS

c MODE-LOCKED LASER: ANALYSIS ON THE BASIS OF

c THE MOMENTUM METHOD

c

c \ \qquad \qquad \qquad \qquad \qquad V.L.Kalashnikov

c\qquad \qquad Photonics Institute, Technical University of Vienna

c\qquad \qquad \qquad \qquad \qquad e-mail: kalashnikov@tuwien.ac.at\qquad
\qquad \qquad \qquad

c\qquad \qquad \qquad web-site: http://www.geocities.com/optomaplev\qquad
\qquad \qquad

cccccccccccccccccccccccccccccccccccccccccccccccccccccccccccccccccccccccccc

c

c This program is realized on the basis of the computer algebra approach

c (the corresponding description can be found on http://www.geocities.com/

c optomaplev/programs/fortran.html)

c

c PARAMETERS:

c

c a is the distance from the out-put plane mirror to the first folding

c mirror;

c c is the distance from the totally reflecting plane mirror to the

c second folding mirror;

c b is the distance between folding mirrors;

c bs and bf give the limits of the scanning on b;

c b1 is the distance of the active medium facet from the first folding

c mirror;

c b1s is the starting b1

c f is the focus length of the folding mirrors;

c z is the active medium length;

c n is the refractive index of the active medium;

c lambda is the generation wavelength;

c eps1 gives the criterion of the convergence to the steady-state

c solution;

c P is the pump power in watts;

c wp is the pump beam size;

c l is the loss coefficient on the pump wavelength (in 1/cm);

c loss is the out-put loss coefficient;

c am is the maximum gain coefficient;

c Pcr is the critical power of the self-focusing in the active medium

c (in watts);

c Dam is the group-velocity dispersion of the active medium (in fs\symbol{94}%
2/cm,

c normal dispersion has a negative sign);

c Is is the gain saturation intensity;

c\qquad

c q is the complex Gaussian beam parameter;

c Pw is the pulse power;

c delta is the pulse width (for Gaussian pulse);

c psi is the pulse chirp;

c alpha and beta is the generation beam parameters (see description)

c

c Lengths are given in centimeters; powers are given in watts

c

c

\qquad COMPLEX*16 i,qs,q(15),qq,t \ ! i is the imaginary unit, qs is the
initial q

\qquad REAL*8 a,c,b1,b2,b,f,z,eps1,eps2,ro(15),n,x,ka(4),qu(4),tau(4)

\qquad REAL*8 delta(4),psi(4),del,delta0,ps,psi0,tg

\qquad REAL*8 lambda,Pi,b1s,bs,bf,beta0,alpha0,k,beta,alpha,kappa,wp,P,l

\qquad REAL*8 gamma,gamma0,loss,Pw,g,Is,S,am,Pcr,Pd(1),Der,Dam,Lcav,Tcav

\qquad INTEGER Num

\qquad DATA a,c,z,f,n,eps1/30d0,60d0,0.28d0,15d0,2.442d0,1d-3/

\qquad DATA b1s,bs,bf,wp,P,l,loss/0d0,20d0,55d0,100d-4,1.5d0,6d0,5d-2/

\qquad DATA Pi,lambda/3.14159265358979323846264338d0,2.5d-4/

\qquad DATA am,Pcr,cv,Dam,Is/9d0,965d3,3d10,-2054d0,1d4/

\qquad i=dcmplx(0.,1.)\qquad

\qquad open(1,file='ro.dat')

\qquad open(2,file='der.dat')

\qquad write(1,*)a,c,wp,P

\qquad ! w0=100 mkm is the initial size of the plane wave:

\qquad qs = (0.,.7957747152)

\qquad ! wmax=10 cm is maximum size of the simulated mode

\qquad eps2 = 1d2*Pi/lambda

\qquad k = 2d0*Pi/lambda ! is the wave number

\qquad ! It takes into account the pump wave damping in the active medium:

\qquad kappa =.5506035739d0*P

\qquad \qquad \qquad \qquad \qquad \qquad \qquad

\qquad S = 2d0*0.48d0*P*.2900302115d-4/Pi/wp**2 ! is the pump parameter

\qquad \qquad \qquad \qquad ! sigma\_a*T\_r/h/nu\_p=.2900302115d-4.

\qquad \qquad \qquad \qquad ! 0.48 - averaging along propagation axis.

\qquad \qquad \qquad \qquad ! sigma\_a, T\_r, nu\_p are the absorption

\qquad \qquad \qquad \qquad ! cross section of the active medium,

\qquad \qquad \qquad \qquad ! the gain relaxation time, and the gain

\qquad \qquad \qquad \qquad ! wavelength, respectively

\qquad x = z/n\qquad ! is the optical length of the gain medium

\qquad tg = 4d0/z\qquad ! is the gain band width in fs/cm

\qquad dx = x/1d3\qquad ! is the step size

\qquad DO I1=1,201\qquad \qquad \qquad \qquad \qquad \qquad ! scanning on b

\qquad b = bs + (bf-bs)*(I1-1)/200d0

\qquad b1f = b-z

\qquad DO I2=1,201\qquad \qquad \qquad \qquad \qquad \qquad ! scanning on b1

\qquad b1 = b1s + (b1f-b1s)*(I2-1)/200d0

\qquad b2 = b-b1-z

\qquad Lcav = a + b + c ! is the cavity length

\qquad Tcav = (2d0*Lcav/cv)*1d15\qquad ! is the cavity period fs

\qquad write(*,*)I1,I2 ! just step numbers !

\qquad gamma0 = 0d0

\qquad Pw = 10d0\qquad \qquad ! initial pulse power in watts

\qquad

\qquad delta0 = 1d3\qquad ! initial pulse width in fs

\qquad psi0 = 0d0\qquad \qquad ! initial pulse chirp

c\qquad beam propagation

\qquad Num = 0

\qquad qq = qs

2\qquad continue

\qquad Num = Num + 1

\qquad Pwold = Pw

c\qquad ABCD-modul: output mirror - folding mirror - active medium

\qquad q(1) = qq

\qquad q(2) = q(1) + a

\qquad q(3) = 1d0/(1d0/q(2) - 1d0/f)

\qquad q(4) = q(3) + b1

c---------------------------------------------------------------

\qquad if(-1d0/dimag(1d0/q(4)).lt.0d0)goto 1

\qquad beta0 = -k*dreal(1d0/q(4))/2d0

\qquad alpha0 = dsqrt(-1d0/dimag(1d0/q(4))/k)

\qquad g = am*S/(1d0 + S +

\#Pw*delta0*dsqrt(Pi/2d0)/(Pi*alpha0**2*Is*Tcav))\qquad ! saturated gain

\qquad

c\qquad Active medium (Runge-Kutta fourth-order method)

\qquad

\qquad DO J=1,1000

\qquad ka(1) = (-0.5d0/alpha0**4 + 2d0*beta0**2 +

\#2d0*k*kappa*dexp(-l*dx*(J-1))/wp**2)/k +

\#(dsqrt(2d0)/Pi)*(Pw/Pcr)*dexp(2*gamma0)/alpha0**4/k\qquad \qquad \qquad
\qquad !beta

\qquad qu(1) = - 2d0*beta0*alpha0/k -

\#2d0*g*alpha0**3/(1d0+2d0*alpha0**2/wp**2)**(3/2)/wp**2 !alpha

\qquad tau(1) = g/(1d0 + 2d0*alpha0**2/wp**2) - 2d0*tg**2/delta0**2 + !gamma

\#Dam*psi0

\qquad delta(1) = 2d0*tg**2*(1d0/delta0 - delta0**3*psi0**2) -\qquad \qquad
\qquad !delta

\#2d0*Dam*delta0*psi0

\qquad psi(1) = 2d0*Dam*(psi0**2 - 1d0/delta0**4) - 8d0*tg**2*psi0/ !psi

\#delta0**2 + (2d0/Pi)*(Pw/Pcr)*dexp(2*gamma0)/alpha0**2/delta0**2\qquad

\qquad

\qquad

\qquad ka(2) = (-0.5d0/(alpha0+0.5d0*dx*qu(1))**4 +

\#2d0*(beta0+0.5d0*dx*ka(1))**2 +

\#2d0*k*kappa*dexp(-l*dx*(J-0.5d0))/wp**2)/k +

\#(dsqrt(2d0)/Pi)*(Pw/Pcr)*dexp(2*gamma0+0.5d0*dx*tau(1))/

\#(alpha0+0.5d0*dx*qu(1))**4/k

\qquad qu(2) = - 2d0*(beta0+0.5d0*dx*ka(1))*(alpha0+0.5d0*dx*qu(1))/k -

\#2d0*g*(alpha0+0.5d0*dx*qu(1))**3/(1d0+

\#2d0*(alpha0+0.5d0*dx*qu(1))**2/wp**2)**(3/2)/wp**2

\qquad tau(2) = g/(1d0 + 2d0*(alpha0+0.5d0*dx*qu(1))**2/wp**2) -

\#2d0*tg**2/(delta0+0.5d0*dx*delta(1))**2+Dam*(psi0+0.5d0*dx*psi(1))

\qquad delta(2) = 2d0*tg**2*(1d0/(delta0+0.5d0*dx*delta(1)) -

\#(delta0+0.5d0*dx*delta(1))**3*(psi0+0.5d0*dx*psi(1))**2) -

\#2d0*Dam*(delta0+0.5d0*dx*delta(1))*(psi0+0.5d0*dx*psi(1))

\qquad psi(2) = 2d0*Dam*((psi0+0.5d0*dx*psi(1))**2 -

\#1d0/(delta0+0.5d0*dx*delta(1))**4) -

\#8d0*tg**2*(psi0+0.5d0*dx*psi(1))/

\#(delta0+0.5d0*dx*delta(1))**2 + (2d0/Pi)*(Pw/Pcr)*

\#dexp(2*(gamma0+0.5d0*dx*tau(1)))/(alpha0+0.5d0*dx*qu(1))**2/

\#(delta0+0.5d0*dx*delta(1))**2\qquad

\qquad ka(3) = (-0.5d0/(alpha0+0.5d0*dx*qu(2))**4 +

\#2d0*(beta0+0.5d0*dx*ka(2))**2 +

\#2d0*k*kappa*dexp(-l*dx*(J-0.5d0))/wp**2)/k +

\#(dsqrt(2d0)/Pi)*(Pw/Pcr)*dexp(2*gamma0+0.5d0*dx*tau(2))/

\#(alpha0+0.5d0*dx*qu(2))**4/k

\qquad qu(3) = - 2d0*(beta0+0.5d0*dx*ka(2))*(alpha0+0.5d0*dx*qu(2))/k -

\#2d0*g*(alpha0+0.5d0*dx*qu(2))**3/(1d0+

\#2d0*(alpha0+0.5d0*dx*qu(2))**2/wp**2)**(3/2)/wp**2

\qquad tau(3) = g/(1d0 + 2d0*(alpha0+0.5d0*dx*qu(2))**2/wp**2) -

\#2d0*tg**2/(delta0+0.5d0*dx*delta(2))**2+Dam*(psi0+0.5d0*dx*psi(2))

\qquad delta(3) = 2d0*tg**2*(1d0/(delta0+0.5d0*dx*delta(2)) -

\#(delta0+0.5d0*dx*delta(2))**3*(psi0+0.5d0*dx*psi(2))**2) -

\#2d0*Dam*(delta0+0.5d0*dx*delta(2))*(psi0+0.5d0*dx*psi(2))

\qquad psi(3) = 2d0*Dam*((psi0+0.5d0*dx*psi(2))**2 -

\#1d0/(delta0+0.5d0*dx*delta(2))**4) -

\#8d0*tg**2*(psi0+0.5d0*dx*psi(2))/

\#(delta0+0.5d0*dx*delta(2))**2 + (2d0/Pi)*(Pw/Pcr)*

\#dexp(2*(gamma0+0.5d0*dx*tau(2)))/(alpha0+0.5d0*dx*qu(2))**2/

\#(delta0+0.5d0*dx*delta(2))**2\qquad

\qquad ka(4) = (-0.5d0/(alpha0+dx*qu(3))**4 +

\#2d0*(beta0+dx*ka(3))**2 +

\#2d0*k*kappa*dexp(-l*dx*J)/wp**2)/k +

\#(dsqrt(2d0)/Pi)*(Pw/Pcr)*dexp(2*gamma0+dx*tau(3))/

\#(alpha0+dx*qu(3))**4/k

\qquad qu(4) = - 2d0*(beta0+dx*ka(3))*(alpha0+dx*qu(3))/k -

\#2d0*g*(alpha0+dx*qu(3))**3/(1d0+

\#2d0*(alpha0+dx*qu(3))**2/wp**2)**(3/2)/wp**2

\qquad tau(4) = g/(1d0 + 2d0*(alpha0+dx*qu(3))**2/wp**2) -

\#2d0*tg**2/(delta0+dx*delta(3))**2 + Dam*(psi0+dx*psi(3))

\qquad delta(4) = 2d0*tg**2*(1d0/(delta0+dx*delta(3)) -

\#(delta0+dx*delta(3))**3*(psi0+dx*psi(3))**2) -

\#2d0*Dam*(delta0+dx*delta(3))*(psi0+dx*psi(3))

\qquad psi(4) = 2d0*Dam*((psi0+dx*psi(3))**2 -

\#1d0/(delta0+dx*delta(3))**4) -

\#8d0*tg**2*(psi0+dx*psi(3))/

\#(delta0+dx*delta(3))**2 + (2d0/Pi)*(Pw/Pcr)*

\#dexp(2*(gamma0+dx*tau(3)))/(alpha0+dx*qu(3))**2/

\#(delta0+dx*delta(3))**2\qquad

\qquad beta = beta0 + (dx/6d0)*(ka(1)+2d0*ka(2)+2d0*ka(3)+ka(4))

\qquad alpha = alpha0 + (dx/6d0)*(qu(1)+2d0*qu(2)+2d0*qu(3)+qu(4))

\qquad gamma = gamma0 + (dx/6d0)*(tau(1)+2d0*tau(2)+2d0*tau(3)+tau(4))

\qquad del=delta0+(dx/6d0)*(delta(1)+2d0*delta(2)+2d0*delta(3)+delta(4))

\qquad ps=psi0+(dx/6d0)*(psi(1)+2d0*psi(2)+2d0*psi(3)+psi(4))

\qquad if(alpha.lt.0.or.alpha.gt.10.or.gamma.lt.0.or.gamma.gt.10.

\#or.del.lt.0.)goto 1

\qquad alpha0 = alpha

\qquad beta0 = beta

\qquad gamma0 = gamma

\qquad delta0 = del

\qquad psi0 = ps

\qquad

\qquad END DO

\qquad Pw = Pw*dexp(2d0*gamma0)

\qquad \qquad gamma0 = 0d0

\qquad if(Pw.le.1d-10.or.Pw.gt.1d20)goto 1

\qquad t = k/(-2d0*beta0 - i/alpha0**2)

\qquad q(5) = dcmplx(dreal(t),dimag(t))

c-----------------------------------------------------------

c ABCD-modul: active medium - second folding mirror - second flat mirror
(and backwards)

\qquad q(6) = q(5) + b2

\qquad q(7) = 1d0/(1d0/q(6) - 1d0/f)

\qquad q(8) = q(7) + c

\qquad q(9) = q(8) + c

\qquad q(10) = 1d0/(1d0/q(9) - 1d0/f)

\qquad q(11) = q(10) + b2

\qquad

c-----------------------------------------------------------

\qquad if(-1d0/dimag(1d0/q(11)).lt.0d0)goto 1

\qquad beta0 = -k*dreal(1d0/q(11))/2d0

\qquad alpha0 = dsqrt(-1d0/dimag(1d0/q(11))/k)

\qquad

\qquad g = am*S/(1d0 + S +

\#Pw*delta0*dsqrt(Pi/2d0)/(Pi*alpha0**2*Is*Tcav))

\qquad

c\qquad Active medium (Runge-Kutta fourth-order method)

\qquad DO J=1,1000

\qquad ka(1) = (-0.5d0/alpha0**4 + 2d0*beta0**2 +

\#2d0*k*kappa*dexp(-l*dx*(J-1))/wp**2)/k +

\#(dsqrt(2d0)/Pi)*(Pw/Pcr)*dexp(2*gamma0)/alpha0**4/k

\qquad qu(1) = - 2d0*beta0*alpha0/k -

\#2d0*g*alpha0**3/(1d0+2d0*alpha0**2/wp**2)**(3/2)/wp**2

\qquad tau(1) = g/(1d0 + 2d0*alpha0**2/wp**2) - 2d0*tg**2/delta0**2 +

\#Dam*psi0

\qquad delta(1) = 2d0*tg**2*(1d0/delta0 - delta0**3*psi0**2) -

\#2d0*Dam*delta0*psi0

\qquad psi(1) = 2d0*Dam*(psi0**2 - 1d0/delta0**4) - 8d0*tg**2*psi0/

\#delta0**2 + (2d0/Pi)*(Pw/Pcr)*dexp(2*gamma0)/alpha0**2/delta0**2

\qquad

\qquad

\qquad ka(2) = (-0.5d0/(alpha0+0.5d0*dx*qu(1))**4 +

\#2d0*(beta0+0.5d0*dx*ka(1))**2 +

\#2d0*k*kappa*dexp(-l*dx*(J-0.5d0))/wp**2)/k +

\#(dsqrt(2d0)/Pi)*(Pw/Pcr)*dexp(2*gamma0+0.5d0*dx*tau(1))/

\#(alpha0+0.5d0*dx*qu(1))**4/k

\qquad qu(2) = - 2d0*(beta0+0.5d0*dx*ka(1))*(alpha0+0.5d0*dx*qu(1))/k -

\#2d0*g*(alpha0+0.5d0*dx*qu(1))**3/(1d0+

\#2d0*(alpha0+0.5d0*dx*qu(1))**2/wp**2)**(3/2)/wp**2

\qquad tau(2) = g/(1d0 + 2d0*(alpha0+0.5d0*dx*qu(1))**2/wp**2) -

\#2d0*tg**2/(delta0+0.5d0*dx*delta(1))**2+Dam*(psi0+0.5d0*dx*psi(1))

\qquad delta(2) = 2d0*tg**2*(1d0/(delta0+0.5d0*dx*delta(1)) -

\#(delta0+0.5d0*dx*delta(1))**3*(psi0+0.5d0*dx*psi(1))**2) -

\#2d0*Dam*(delta0+0.5d0*dx*delta(1))*(psi0+0.5d0*dx*psi(1))

\qquad psi(2) = 2d0*Dam*((psi0+0.5d0*dx*psi(1))**2 -

\#1d0/(delta0+0.5d0*dx*delta(1))**4) -

\#8d0*tg**2*(psi0+0.5d0*dx*psi(1))/

\#(delta0+0.5d0*dx*delta(1))**2 + (2d0/Pi)*(Pw/Pcr)*

\#dexp(2*(gamma0+0.5d0*dx*tau(1)))/(alpha0+0.5d0*dx*qu(1))**2/

\#(delta0+0.5d0*dx*delta(1))**2\qquad

\qquad ka(3) = (-0.5d0/(alpha0+0.5d0*dx*qu(2))**4 +

\#2d0*(beta0+0.5d0*dx*ka(2))**2 +

\#2d0*k*kappa*dexp(-l*dx*(J-0.5d0))/wp**2)/k +

\#(dsqrt(2d0)/Pi)*(Pw/Pcr)*dexp(2*gamma0+0.5d0*dx*tau(2))/

\#(alpha0+0.5d0*dx*qu(2))**4/k

\qquad qu(3) = - 2d0*(beta0+0.5d0*dx*ka(2))*(alpha0+0.5d0*dx*qu(2))/k -

\#2d0*g*(alpha0+0.5d0*dx*qu(2))**3/(1d0+

\#2d0*(alpha0+0.5d0*dx*qu(2))**2/wp**2)**(3/2)/wp**2

\qquad tau(3) = g/(1d0 + 2d0*(alpha0+0.5d0*dx*qu(2))**2/wp**2) -

\#2d0*tg**2/(delta0+0.5d0*dx*delta(2))**2+Dam*(psi0+0.5d0*dx*psi(2))

\qquad delta(3) = 2d0*tg**2*(1d0/(delta0+0.5d0*dx*delta(2)) -

\#(delta0+0.5d0*dx*delta(2))**3*(psi0+0.5d0*dx*psi(2))**2) -

\#2d0*Dam*(delta0+0.5d0*dx*delta(2))*(psi0+0.5d0*dx*psi(2))

\qquad psi(3) = 2d0*Dam*((psi0+0.5d0*dx*psi(2))**2 -

\#1d0/(delta0+0.5d0*dx*delta(2))**4) -

\#8d0*tg**2*(psi0+0.5d0*dx*psi(2))/

\#(delta0+0.5d0*dx*delta(2))**2 + (2d0/Pi)*(Pw/Pcr)*

\#dexp(2*(gamma0+0.5d0*dx*tau(2)))/(alpha0+0.5d0*dx*qu(2))**2/

\#(delta0+0.5d0*dx*delta(2))**2\qquad

\qquad ka(4) = (-0.5d0/(alpha0+dx*qu(3))**4 +

\#2d0*(beta0+dx*ka(3))**2 +

\#2d0*k*kappa*dexp(-l*dx*J)/wp**2)/k +

\#(dsqrt(2d0)/Pi)*(Pw/Pcr)*dexp(2*gamma0+dx*tau(3))/

\#(alpha0+dx*qu(3))**4/k

\qquad qu(4) = - 2d0*(beta0+dx*ka(3))*(alpha0+dx*qu(3))/k -

\#2d0*g*(alpha0+dx*qu(3))**3/(1d0+

\#2d0*(alpha0+dx*qu(3))**2/wp**2)**(3/2)/wp**2

\qquad tau(4) = g/(1d0 + 2d0*(alpha0+dx*qu(3))**2/wp**2) -

\#2d0*tg**2/(delta0+dx*delta(3))**2 + Dam*(psi0+dx*psi(3))

\qquad delta(4) = 2d0*tg**2*(1d0/(delta0+dx*delta(3)) -

\#(delta0+dx*delta(3))**3*(psi0+dx*psi(3))**2) -

\#2d0*Dam*(delta0+dx*delta(3))*(psi0+dx*psi(3))

\qquad psi(4) = 2d0*Dam*((psi0+dx*psi(3))**2 -

\#1d0/(delta0+dx*delta(3))**4) -

\#8d0*tg**2*(psi0+dx*psi(3))/

\#(delta0+dx*delta(3))**2 + (2d0/Pi)*(Pw/Pcr)*

\#dexp(2*(gamma0+dx*tau(3)))/(alpha0+dx*qu(3))**2/

\#(delta0+dx*delta(3))**2\qquad

\qquad beta = beta0 + (dx/6d0)*(ka(1)+2d0*ka(2)+2d0*ka(3)+ka(4))

\qquad alpha = alpha0 + (dx/6d0)*(qu(1)+2d0*qu(2)+2d0*qu(3)+qu(4))

\qquad gamma = gamma0 + (dx/6d0)*(tau(1)+2d0*tau(2)+2d0*tau(3)+tau(4))

\qquad del=delta0+(dx/6d0)*(delta(1)+2d0*delta(2)+2d0*delta(3)+delta(4))

\qquad ps=psi0+(dx/6d0)*(psi(1)+2d0*psi(2)+2d0*psi(3)+psi(4))

\qquad if(alpha.lt.0.or.alpha.gt.10.or.gamma.lt.0.or.gamma.gt.10.

\#or.del.lt.0.)goto 1

\qquad alpha0 = alpha

\qquad beta0 = beta

\qquad gamma0 = gamma

\qquad delta0 = del

\qquad psi0 = ps

\qquad

\qquad END DO

c-------------------------------------------------------------

\qquad Pw = Pw*dexp(2d0*gamma0)

\qquad \qquad gamma0 = 0d0

\qquad if(Pw.le.1d-10.or.Pw.gt.1d20)goto 1

\qquad t = k/(-2d0*beta0 - i/alpha0**2)

\qquad q(12) = dcmplx(dreal(t),dimag(t))\qquad

c\qquad ABCD-modul for the residuary propagation up to out-put mirror

\qquad q(13) = q(12) + b1

\qquad q(14) = 1d0/(1d0/q(13) - 1d0/f)

\qquad q(15) = q(14) + a

\qquad Pw = Pw*dexp(-loss)

c-------------------------------------------------------------

\qquad if(Pw.le.1d-10.or.Pw.gt.1d20)goto 1\qquad ! criteria for the power

\qquad if(Num.gt.5000)goto 1\qquad \qquad \qquad \qquad ! and iteration
number

\qquad

c\qquad w\symbol{94}2*Pi/lambda converts the initial part of the beam
parameter to the beam size

\qquad

\qquad DO I4=1,15

\qquad ro(I4) = -1d0/dimag(1d0/q(I4))

\qquad END DO

c\qquad Beam is to have the positive size and hasn't to be too large

\qquad if(ro(1).le.0..or.ro(2).le.0..or.ro(3).le.0..or.

2\qquad ro(4).le.0..or.ro(5).le.0..or.ro(6).le.0..or.

3\qquad ro(7).le.0..or.ro(8).le.0..or.ro(9).le.0..or.

4\qquad ro(10).le.0..or.ro(11).le.0..or.ro(12).le.0..or.

5\qquad ro(13).le.0..or.ro(14).le.0..or.ro(15).le.0.)goto 1

\qquad if(ro(1).gt.eps2.or.ro(2).gt.eps2.or.ro(3).gt.eps2.or.

2\qquad ro(4).gt.eps2.or.ro(5).gt.eps2.or.ro(6).gt.eps2.or.

3\qquad ro(7).gt.eps2.or.ro(8).gt.eps2.or.ro(9).gt.eps2.or.

4\qquad ro(10).gt.eps2.or.ro(11).gt.eps2.or.ro(12).gt.eps2.or.

5\qquad ro(13).gt.eps2.or.ro(14).gt.eps2.or.ro(15).gt.eps2)goto 1

\qquad qq = q(15)

c\qquad Pulse power stability

\qquad if(abs(ro(15)/ro(1)-1d0).gt.eps1.

\#or.abs(Pw/Pwold-1d0).gt.eps1)goto 2

c\qquad Out-put for the stable pulse

\qquad write(1,*)b,b1,Pw,sqrt(ro(5)*lambda/Pi),delta0,Num

1\qquad continue

\qquad \qquad \qquad END DO

\qquad \qquad \qquad END DO

\qquad close(1)

\qquad close(2)

\qquad STOP

\qquad END

\qquad

\end{document}